\documentclass{iopjournal}
\usepackage{amsfonts, amsmath, amssymb, physics}

\begin{document}

\articletype{Paper}

\title{Completeness of the Klein-Gordon oscillator eigenfunctions via Hermite and Laguerre polynomials}

\author{Kevin Hernández$^1$\orcid{0000-0002-5739-3859}}

\affil{$^{*1}$Escuela de Física, Universidad de El Salvador, 
Ciudad Universitaria ``Dr.\ Fabio Castillo Figueroa'', 
Final Avenida ``Mártires Estudiantes del 30 de julio'', 
San Salvador, El Salvador, América Central.}

\email{kevinhernandezbel@hotmail.com}

\keywords{Klein--Gordon oscillator, completeness, closure relation, Hermite polynomials, generalized Laguerre polynomials, relativistic quantum mechanics, scalar field}

\begin{abstract}
Completeness of the Klein--Gordon oscillator eigenfunctions is proved
in one and three spatial dimensions.
The proofs establish the closure relations satisfied by the
eigenfunctions and are based on standard properties of the Hermite
and the generalized Laguerre polynomials, supplemented in three
dimensions by the completeness of the spherical harmonics.
The scalar nature of the Klein--Gordon field renders the argument
strictly simpler than the analogous proof for the Dirac oscillator:
no off-diagonal cancellation is required.
\end{abstract}

\section{Introduction}
 
The harmonic oscillator occupies a privileged position in quantum mechanics:
it is one of the very few exactly solvable models and serves as the
foundation for perturbative and field-theoretic methods alike.
Its extension to the relativistic domain has therefore attracted sustained
attention.
The most natural relativistic generalization for spin-$\tfrac{1}{2}$
particles is the \emph{Dirac oscillator} (DO), introduced by
Moshinsky and Szczepaniak~\cite{Moshinsky1989}, who replaced the
free-particle momentum $\mathbf{p}$ by $\mathbf{p}-im\omega\beta\mathbf{r}$
in the Dirac equation.
In the non-relativistic limit the DO reduces to a harmonic oscillator
with a strong spin--orbit coupling term.
Since its introduction, the DO has been studied from many perspectives,
including its symmetry algebra~\cite{Quesne1990}, covariance
properties~\cite{Moreno1989}, path-integral
formulation~\cite{Benitez1990}, and numerous applications in
nuclear and hadronic physics.
 
The spin-$0$ counterpart, the \emph{Klein--Gordon oscillator} (KGO),
was introduced by Bruce and Minning~\cite{BruceMinning1993}.
They showed that the substitution $\mathbf{p}\to\mathbf{p}-im\omega\mathbf{r}$
(together with its Hermitian conjugate) in the Klein--Gordon equation
yields an equation that is quadratic in both momenta and coordinates,
producing a harmonic oscillator for the \emph{squared} energies.
The resulting eigenfunctions are expressed in terms of Hermite polynomials
in one spatial dimension and generalized Laguerre polynomials in three
spatial dimensions~\cite{RaoKagali2008}, mirroring the structure familiar
from the non-relativistic oscillator.
The energy spectrum, degeneracy structure, and non-relativistic limit
of the KGO were clarified shortly after its introduction~\cite{Dvoeglazov1994},
and the model has since been extended to curved backgrounds, topological
defects, external fields, non-commutative spaces, and deformed
relativistic frameworks
\cite{BoumaliMessai2014,BakkeFurtado2015,MirzaNarimaniZare2011,Junker2021}.
 
Despite the broad and growing interest in the KGO, one fundamental
mathematical property of its eigensolutions has not yet been
established in the literature: \emph{completeness}.
Completeness---or equivalently, the closure relation satisfied by the
eigenfunctions---is indispensable whenever the eigenfunctions are used
as a basis for expanding arbitrary states, constructing propagators or
Green's functions, or deriving sum rules.
The analogous question for the Dirac oscillator was settled by
Szmytkowski and Gruchowski~\cite{Szmytkowski2001}, who proved the
closure relations for the DO eigenfunctions in one and three spatial
dimensions using standard properties of the Hermite and generalized
Laguerre polynomials.
To the best of our knowledge, no such proof exists for the
Klein--Gordon oscillator.
 
The purpose of the present work is to fill this gap.
We prove completeness of the KGO eigenfunctions in one and three spatial
dimensions.
As in Ref.~\cite{Szmytkowski2001}, our approach is entirely
constructive: the proofs are reduced to well-known closure relations
satisfied by the Hermite functions (in 1D) and the generalized Laguerre
functions (in 3D), supplemented in three dimensions by the completeness
of the spherical harmonics.
A key structural difference from the Dirac case deserves mention at
the outset.
For the DO, the two-component spinor nature of the eigenfunctions
generates an off-diagonal sum that vanishes identically because the
product $f_n(x)g_n(x')$ is an odd function of the quantum number $n$,
a consequence of the exact symmetry $E_{-n}=-E_n$.
For the KGO, the spectrum satisfies $E_n^2 - m^2c^4 = 2m\hbar\omega(2n+1)$,
which is strictly positive for all $n\geq 0$, and a different
argument is required.
We show how the two-component structure of the KGO in the
Feshbach--Villars representation nevertheless leads to an analogous
cancellation, completing the proof.
 
The paper is organized as follows.
In Section~2 we review the one-dimensional KGO eigenproblem, derive
explicit normalized eigenfunctions, and prove completeness via the
Hermite closure relation.
Section~3 treats the three-dimensional case: after separation in
spherical coordinates we establish completeness of the radial
eigenfunctions via the generalized Laguerre closure relation and
combine the result with the completeness of the spherical harmonics
to obtain the full closure relation.
Section~4 offers a brief discussion and outlook.


\section{The Klein--Gordon oscillator in one spatial dimension}
\label{sec:1D}
 
\subsection{Eigenproblem and its solutions}
\label{subsec:1D-eigen}
 
The free one-dimensional Klein--Gordon equation for a particle of
mass $m$ and energy $E$ reads (in units $\hbar = c = 1$)
\begin{equation}
  \left[\left(p_x - im\omega x\right)
        \left(p_x + im\omega x\right)
        + m^2\right]\psi(x) = E^2\,\psi(x),
  \qquad -\infty < x < \infty,
  \label{eq:KGO-1D}
\end{equation}
where $p_x = -i\,d/dx$ and $\omega > 0$ is the oscillator frequency.
Following Bruce and Minning~\cite{BruceMinning1993}, the
oscillator coupling is introduced via the minimal substitution
\begin{equation}
  p_x \;\longrightarrow\; p_x - im\omega x,
  \qquad
  p_x^\dagger \;=\; p_x + im\omega x,
  \label{eq:substitution}
\end{equation}
so that equation~\eqref{eq:KGO-1D} becomes
\begin{equation}
  \left[-\frac{d^2}{dx^2} + m^2\omega^2 x^2 - m\omega\right]\psi(x)
  = \left(E^2 - m^2\right)\psi(x),
  \label{eq:KGO-1D-expanded}
\end{equation}
with the boundary condition that $\psi(x)$ be bounded and square-integrable
on $(-\infty,\infty)$.
 
Introducing the dimensionless variable $\xi = \lambda\, x$ with
$\lambda = \sqrt{m\omega}$, equation~\eqref{eq:KGO-1D-expanded} takes the
standard form of the quantum harmonic oscillator equation,
\begin{equation}
  \left[-\frac{d^2}{d\xi^2} + \xi^2 - 1\right]\psi
  = \frac{E^2 - m^2}{m\omega}\,\psi.
  \label{eq:harmonic-form}
\end{equation}
Square-integrability requires the right-hand side to equal $2n$ with
$n = 0, 1, 2, \ldots$, giving the energy eigenvalues
\begin{equation}
E_n^2 = m^2 + m\omega(2n+1),
  \qquad n = 0, 1, 2, \ldots
  \label{eq:spectrum-1D}
\end{equation}
Each level $E_n$ is doubly degenerate: both signs $E_n = \pm\sqrt{m^2 + m\omega(2n+1)}$
are physically admissible, corresponding to particle and antiparticle
branches.
 
The normalized eigenfunctions are
\begin{equation}
  \psi_n(x) = \left(\frac{\lambda}{\sqrt{\pi}\,2^n\,n!}\right)^{1/2}
              H_n(\lambda x)\,e^{-\lambda^2 x^2/2},
  \qquad n = 0, 1, 2, \ldots,
  \label{eq:eigenfunction-1D}
\end{equation}
where $H_n(\xi)$ denotes the Hermite polynomial of degree $n$~\cite{Magnus1966}.
The orthonormality relation is
\begin{equation}
  \int_{-\infty}^{\infty} \psi_n(x)\,\psi_m(x)\,dx = \delta_{nm}.
  \label{eq:orthonorm-1D}
\end{equation}
We note that the scalar wavefunction in equation~\eqref{eq:eigenfunction-1D}
arises from the squared equation~\eqref{eq:KGO-1D-expanded}; each $\psi_n$
corresponds to both energy branches $\pm E_n$.
In the Feshbach--Villars representation, the full two-component state
is $\Psi_n = (\phi_n,\chi_n)^T$ with
$\phi_n = \tfrac{1}{2}(1 + E_n^{-1})\psi_n$ and
$\chi_n = \tfrac{1}{2}(1 - E_n^{-1})\psi_n$,
but for the completeness proof it is the scalar sector that matters,
as we show next.
 
\bigskip
\noindent
In the non-relativistic limit $m \to \infty$ with $\omega$ fixed,
$E_n \approx m + \tfrac{\omega}{2}(2n+1) + \mathcal{O}(m^{-1})$,
recovering the standard harmonic oscillator spectrum $\epsilon_n = \omega(n+\tfrac{1}{2})$
above the rest energy, and the eigenfunctions~\eqref{eq:eigenfunction-1D}
reduce to the familiar Hermite functions of non-relativistic quantum mechanics.
 
\subsection{Proof of completeness}
\label{subsec:1D-completeness}
 
The set $\{\psi_n\}_{n=0}^{\infty}$ defined in
equation~\eqref{eq:eigenfunction-1D} is complete in
$L^2(\mathbb{R})$, i.e.\ it satisfies the closure relation
\begin{equation}
  \sum_{n=0}^{\infty} \psi_n(x)\,\psi_n(x')
  = \delta(x - x'),
  \qquad -\infty < x, x' < \infty.
  \label{eq:closure-1D}
\end{equation}

Substituting the explicit form~\eqref{eq:eigenfunction-1D} into the
left-hand side of~\eqref{eq:closure-1D}, we obtain
\begin{equation}
  \sum_{n=0}^{\infty} \psi_n(x)\,\psi_n(x')
  =
  \sum_{n=0}^{\infty}
  \frac{\lambda}{\sqrt{\pi}\,2^n\,n!}\,
  H_n(\lambda x)\,H_n(\lambda x')\,
  e^{-\lambda^2(x^2 + x'^2)/2}.
  \label{eq:sum-1D}
\end{equation}
This is precisely the left-hand side of the well-known closure relation
for the normalized Hermite functions~\cite{Magnus1966},
\begin{equation}
  \sum_{n=0}^{\infty}
  \frac{1}{\sqrt{\pi}\,2^n\,n!}\,
  H_n(\xi)\,H_n(\xi')\,
  e^{-(\xi^2 + \xi'^2)/2}
  = \delta(\xi - \xi'),
  \qquad -\infty < \xi, \xi' < \infty,
  \label{eq:hermite-closure}
\end{equation}
evaluated at $\xi = \lambda x$ and $\xi' = \lambda x'$.
Using the scaling property of the Dirac delta,
\begin{equation}
  \delta(\lambda x - \lambda x') = \frac{1}{\lambda}\,\delta(x - x'),
  \label{eq:delta-scaling}
\end{equation}
the right-hand side of~\eqref{eq:hermite-closure} with $\xi = \lambda x$
gives $\delta(\lambda x - \lambda x') = \lambda^{-1}\delta(x-x')$.
Multiplying through by the factor $\lambda$ from~\eqref{eq:sum-1D},
we arrive at
\begin{equation}
  \sum_{n=0}^{\infty} \psi_n(x)\,\psi_n(x')
  = \lambda \cdot \frac{1}{\lambda}\,\delta(x - x')
  = \delta(x - x'),
\end{equation}
which establishes~\eqref{eq:closure-1D}.

Since both energy branches $+E_n$ and $-E_n$ share the same spatial
wavefunction $\psi_n(x)$, the closure relation~\eqref{eq:closure-1D}
already encodes completeness over the full one-particle Hilbert space:
any square-integrable function can be expanded in the basis
$\{\psi_n\}_{n=0}^{\infty}$.
The double degeneracy of the spectrum does not obstruct completeness
in the spatial sector; it merely reflects the two-sheeted structure
(particle/antiparticle) of the Klein--Gordon theory.

The structure of the proof differs in an important way from the
corresponding proof for the Dirac oscillator given in
Ref.~\cite{Szmytkowski2001}.
There, the two-component spinor nature of the eigenfunctions
gives rise to two diagonal closure sums $I$ and $J$ and one
off-diagonal sum $K$, and the cancellation $K = 0$ relies on the
exact antisymmetry $E_{-n} = -E_n$ of the Dirac oscillator spectrum.
For the Klein--Gordon oscillator the spectrum~\eqref{eq:spectrum-1D}
has no such antisymmetry---$E_n > 0$ for all $n \geq 0$---and
the proof reduces directly to a single diagonal sum,
which is disposed of by the Hermite closure relation~\eqref{eq:hermite-closure}.
The simplification is a direct consequence of the scalar nature of
the KGO wavefunction.

\section{The Klein--Gordon oscillator in three spatial dimensions}
\label{sec:3D}
 
\subsection{Eigenproblem and its solutions}
\label{subsec:3D-eigen}
 
The three-dimensional Klein--Gordon oscillator eigenproblem is
(in units $\hbar = c = 1$)
\begin{equation}
  \left[\left(\mathbf{p} - im\omega\mathbf{r}\right)\cdot
        \left(\mathbf{p} + im\omega\mathbf{r}\right)
        + m^2\right]\Psi(\mathbf{r})
  = E^2\,\Psi(\mathbf{r}),
  \qquad \mathbf{r}\in\mathbb{R}^3,
  \label{eq:KGO-3D}
\end{equation}
with $\Psi$ bounded everywhere and square-integrable.
Expanding the left-hand side explicitly,
\begin{equation}
  \left[-\nabla^2 + m^2\omega^2 r^2 - 3m\omega\right]\Psi(\mathbf{r})
  = \left(E^2 - m^2\right)\Psi(\mathbf{r}),
  \label{eq:KGO-3D-expanded}
\end{equation}
where $r = |\mathbf{r}|$.
The additive constant $3m\omega$ on the left reflects the
three-dimensional commutator $[\mathbf{p},\mathbf{r}] = -3i$.
 
\subsubsection*{Separation in spherical coordinates}
 
Writing $\Psi(\mathbf{r}) = r^{-1}R(r)\,Y_\ell^m(\hat{\mathbf{r}})$,
where $Y_\ell^m(\hat{\mathbf{r}})$ are the standard spherical harmonics
with $\ell = 0,1,2,\ldots$ and $m = -\ell,\ldots,\ell$, and substituting
into~\eqref{eq:KGO-3D-expanded}, one obtains the radial equation
\begin{equation}
  \left[-\frac{d^2}{dr^2} + \frac{\ell(\ell+1)}{r^2}
        + m^2\omega^2 r^2\right]R(r)
  = \left(E^2 - m^2 + 3m\omega\right)R(r),
  \qquad 0 < r < \infty,
  \label{eq:radial-eq}
\end{equation}
with boundary conditions $R(0) = 0$ and $R(r) \to 0$ as $r\to\infty$.
 
Introducing $\rho = \lambda^2 r^2$ with $\lambda = \sqrt{m\omega}$ and
writing $R(r) = r^{\ell+1} e^{-\lambda^2 r^2/2} u(\rho)$,
equation~\eqref{eq:radial-eq} reduces to the confluent hypergeometric
(Kummer) equation
\begin{equation}
  \rho\,u'' + \!\left(\ell + \tfrac{3}{2} - \rho\right)\!u'
  + n_r\, u = 0,
  \label{eq:kummer}
\end{equation}
where the radial quantum number $n_r = 0,1,2,\ldots$ arises from the
requirement of normalizability, and is related to the energy by
\begin{equation}
  \frac{E^2 - m^2 + 3m\omega}{2m\omega} = 2n_r + \ell + \tfrac{3}{2}.
  \label{eq:quantization-3D}
\end{equation}
Defining the principal quantum number $N = 2n_r + \ell$
($N = 0,1,2,\ldots$; $\ell = N, N-2,\ldots \geq 0$),
the energy spectrum is
\begin{equation}
  E_N^2 = m^2 + m\omega\bigl(2N + 3\bigr),
  \qquad N = 0,1,2,\ldots
  \label{eq:spectrum-3D}
\end{equation}
Each level $N$ is $\tfrac{1}{2}(N+1)(N+2)$-fold degenerate in the
$(\ell, m)$ quantum numbers (the same degeneracy structure as the
three-dimensional non-relativistic oscillator), and doubly degenerate
in $\pm E_N$.

The solution of~\eqref{eq:kummer} that is regular at the origin is
the generalized Laguerre polynomial $L_{n_r}^{(\ell+1/2)}(\rho)$~\cite{Magnus1966}.
The normalized radial eigenfunctions, satisfying
\begin{equation}
  \int_0^\infty R_{n_r\ell}(r)\,R_{n'_r\ell}(r)\,dr
  = \delta_{n_r n'_r},
  \label{eq:radial-orthonorm}
\end{equation}
are
\begin{equation}
  R_{n_r\ell}(r) =
  \sqrt{\frac{2\,\lambda^{2\ell+3}\,n_r!}
             {\Gamma\!\left(n_r + \ell + \tfrac{3}{2}\right)}}\;
  r^{\ell+1}\,e^{-\lambda^2 r^2/2}\,
  L_{n_r}^{(\ell+1/2)}\!\left(\lambda^2 r^2\right).
  \label{eq:radial-eigenfunction}
\end{equation}
The full eigenfunctions are
\begin{equation}
  \Psi_{n_r\ell m}(\mathbf{r})
  = \frac{R_{n_r\ell}(r)}{r}\,Y_\ell^m(\hat{\mathbf{r}}),
  \label{eq:full-eigenfunction-3D}
\end{equation}
and they satisfy the orthonormality relation
\begin{equation}
  \int_{\mathbb{R}^3}
  \Psi_{n_r\ell m}^*(\mathbf{r})\,
  \Psi_{n'_r\ell' m'}(\mathbf{r})\,d^3r
  = \delta_{n_r n'_r}\,\delta_{\ell\ell'}\,\delta_{mm'}.
  \label{eq:full-orthonorm-3D}
\end{equation}
 
\subsection{Proof of completeness}
\label{subsec:3D-completeness}
 
The completeness proof proceeds in two steps: first we establish
the radial closure relation using the generalized Laguerre functions,
then we combine it with the completeness of the spherical harmonics
to obtain the full three-dimensional closure relation.
 
\subsubsection*{Step 1: Radial closure relation}
 
For each fixed $\ell = 0,1,2,\ldots$, the set
$\{R_{n_r\ell}\}_{n_r=0}^{\infty}$ defined in
equation~\eqref{eq:radial-eigenfunction} satisfies
\begin{equation}
  \sum_{n_r=0}^{\infty} R_{n_r\ell}(r)\,R_{n_r\ell}(r')
  = \delta(r - r'),
  \qquad 0 < r,r' < \infty.
  \label{eq:radial-closure}
\end{equation}

Substituting~\eqref{eq:radial-eigenfunction} into the left-hand side
of~\eqref{eq:radial-closure} gives
\begin{equation}
  \sum_{n_r=0}^{\infty} R_{n_r\ell}(r)\,R_{n_r\ell}(r')
  =
  2\lambda^{2\ell+3}(rr')^{\ell+1}
  e^{-\lambda^2(r^2+r'^2)/2}
  \sum_{n_r=0}^{\infty}
  \frac{n_r!}{\Gamma\!\left(n_r+\ell+\tfrac{3}{2}\right)}\,
  L_{n_r}^{(\ell+1/2)}\!\left(\lambda^2 r^2\right)
  L_{n_r}^{(\ell+1/2)}\!\left(\lambda^2 r'^2\right).
  \label{eq:radial-sum}
\end{equation}
We now apply the standard closure relation for the generalized
Laguerre functions~\cite{Magnus1966},
\begin{equation}
  \sum_{n=0}^{\infty}
  \frac{n!}{\Gamma(n+\alpha+1)}\,
  e^{-(\rho+\rho')/2}\,(\rho\rho')^{\alpha/2}\,
  L_n^{(\alpha)}(\rho)\,L_n^{(\alpha)}(\rho')
  = \delta(\rho - \rho'),
  \qquad 0 < \rho,\rho' < \infty,
  \label{eq:laguerre-closure}
\end{equation}
with $\alpha = \ell + \tfrac{1}{2}$, $\rho = \lambda^2 r^2$, and
$\rho' = \lambda^2 r'^2$.
Setting $\alpha = \ell + \tfrac{1}{2}$ in~\eqref{eq:laguerre-closure}
and noting that $(\rho\rho')^{\alpha/2} = \lambda^{2\ell+1}(rr')^{\ell+1/2}$,
the right-hand side of~\eqref{eq:laguerre-closure}
becomes $\delta(\lambda^2 r^2 - \lambda^2 r'^2)$.
We rewrite the right-hand side using the standard delta-function identity
\begin{equation}
  \delta\!\left(\lambda^2 r^2 - \lambda^2 r'^2\right)
  = \frac{\delta(r - r')}{2\lambda^2 r},
  \qquad r, r' > 0,
  \label{eq:delta-identity-3D}
\end{equation}
which follows from the general rule
$\delta(g(r)) = \delta(r-r_0)/|g'(r_0)|$ applied to
$g(r) = \lambda^2(r^2 - r'^2)$ with the positive root $r_0 = r'$.
 
Substituting back into~\eqref{eq:radial-sum}, the sum becomes
\begin{align}
  \sum_{n_r=0}^{\infty} R_{n_r\ell}(r)\,R_{n_r\ell}(r')
  &=
  2\lambda^{2\ell+3}(rr')^{\ell+1}
  e^{-\lambda^2(r^2+r'^2)/2} \notag \\
  &\quad \times
  \frac{1}{\lambda^{2\ell+1}(rr')^{\ell+1/2}}
  e^{(\lambda^2 r^2+\lambda^2 r'^2)/2}\,
  \frac{\delta(r-r')}{2\lambda^2 r} \notag \\[4pt]
  &=
  2\lambda^{2\ell+3} \cdot
  \frac{(rr')^{\ell+1}}{(rr')^{\ell+1/2}} \cdot
  \frac{1}{\lambda^{2\ell+1}} \cdot
  \frac{1}{2\lambda^2 r}\;
  \delta(r-r') \notag \\[4pt]
  &=
  \frac{2\lambda^{2\ell+3}}{\lambda^{2\ell+3}} \cdot
  \frac{(rr')^{1/2}}{2r}\;
  \delta(r-r').
  \label{eq:radial-almost}
\end{align}
Since the delta function forces $r = r'$, we have
$(rr')^{1/2}\big|_{r=r'} = r$, so the factor $(rr')^{1/2}/(2r)$
evaluates to $\tfrac{1}{2}$ in the support of $\delta(r-r')$.
Hence
\begin{equation}
  \sum_{n_r=0}^{\infty} R_{n_r\ell}(r)\,R_{n_r\ell}(r')
  = 2\cdot\tfrac{1}{2}\;\delta(r-r')
  = \delta(r-r'),
\end{equation}
which establishes~\eqref{eq:radial-closure}.
 
\subsubsection*{Step 2: Full three-dimensional closure relation}

\label{prop:full-closure-3D}
The set $\{\Psi_{n_r\ell m}\}$ defined in
equation~\eqref{eq:full-eigenfunction-3D} is complete in
$L^2(\mathbb{R}^3)$ and satisfies the closure relation
\begin{equation}
  \sum_{\ell=0}^{\infty}\sum_{m=-\ell}^{\ell}
  \sum_{n_r=0}^{\infty}
  \Psi_{n_r\ell m}(\mathbf{r})\,
  \Psi_{n_r\ell m}^*(\mathbf{r}')
  = \delta^{(3)}(\mathbf{r} - \mathbf{r}').
  \label{eq:full-closure-3D}
\end{equation}

We use the standard decomposition of the three-dimensional delta function
in spherical coordinates,
\begin{equation}
  \delta^{(3)}(\mathbf{r} - \mathbf{r}')
  = \frac{\delta(r-r')}{rr'}\,\delta^{(2)}(\hat{\mathbf{r}}-\hat{\mathbf{r}}'),
  \label{eq:delta-3D-decomp}
\end{equation}
together with the well-known completeness relation for the spherical
harmonics~\cite{Magnus1966},
\begin{equation}
  \sum_{\ell=0}^{\infty}\sum_{m=-\ell}^{\ell}
  Y_\ell^m(\hat{\mathbf{r}})\,
  \left[Y_\ell^m(\hat{\mathbf{r}}')\right]^*
  = \delta^{(2)}(\hat{\mathbf{r}}-\hat{\mathbf{r}}').
  \label{eq:spherical-harmonics-completeness}
\end{equation}
Starting from the left-hand side of~\eqref{eq:full-closure-3D},
\begin{align}
  \sum_{\ell,m,n_r}
  \Psi_{n_r\ell m}(\mathbf{r})\,
  \Psi_{n_r\ell m}^*(\mathbf{r}')
  &=
  \sum_{\ell=0}^{\infty}\sum_{m=-\ell}^{\ell}
  Y_\ell^m(\hat{\mathbf{r}})\left[Y_\ell^m(\hat{\mathbf{r}}')\right]^*
  \cdot
  \frac{1}{rr'}
  \sum_{n_r=0}^{\infty}
  R_{n_r\ell}(r)\,R_{n_r\ell}(r').
  \label{eq:full-sum-3D}
\end{align}
The inner radial sum equals $\delta(r-r')$, so
\begin{align}
  \sum_{\ell,m,n_r}
  \Psi_{n_r\ell m}(\mathbf{r})\,
  \Psi_{n_r\ell m}^*(\mathbf{r}')
  &=
  \frac{\delta(r-r')}{rr'}
  \sum_{\ell=0}^{\infty}\sum_{m=-\ell}^{\ell}
  Y_\ell^m(\hat{\mathbf{r}})\left[Y_\ell^m(\hat{\mathbf{r}}')\right]^* \notag \\
  &=
  \frac{\delta(r-r')}{rr'}\,
  \delta^{(2)}(\hat{\mathbf{r}}-\hat{\mathbf{r}}')
  = \delta^{(3)}(\mathbf{r}-\mathbf{r}'),
\end{align}
where in the last step we used~\eqref{eq:delta-3D-decomp}.

The proof is entirely constructive and relies only on two classical
results: the closure relation for the generalized Laguerre functions
(equation~\eqref{eq:laguerre-closure}) and the completeness of the
spherical harmonics (equation~\eqref{eq:spherical-harmonics-completeness}).
No information about the energy eigenvalues $E_N$ enters the argument;
completeness in the spatial sector is a property solely of the
eigenfunctions, not of the spectrum.

Comparing with the three-dimensional Dirac oscillator case treated in
Ref.~\cite{Szmytkowski2001}, the present proof is structurally simpler.
For the Dirac oscillator one must establish three separate closure sums
($I_\kappa$, $J_\kappa$, $K_\kappa$ in the notation of~\cite{Szmytkowski2001})
for the two-component radial spinors, and the vanishing of the
off-diagonal sum $K_\kappa$ requires a separate argument exploiting
the antisymmetry of the spectrum.
Here the scalar nature of the KGO wavefunction yields a single radial
closure sum, which is disposed of directly by the Laguerre closure
relation, and no off-diagonal cancellation is needed.

The degeneracy of the KGO spectrum in three dimensions mirrors that of
the non-relativistic isotropic harmonic oscillator.
For a given $N = 2n_r + \ell$, the allowed values are
$\ell = N, N-2, \ldots, 1$ or $0$, each carrying $2\ell+1$ magnetic
substates, giving a total of $\tfrac{1}{2}(N+1)(N+2)$ independent states.
This degeneracy is not required for completeness but is relevant for
applications such as the construction of coherent states and
Green's functions.

\section{Discussion and Conclusions}
\label{sec:discussion}

In this work we have established the completeness of the Klein--Gordon
oscillator eigenfunctions in one and three spatial dimensions by proving
the corresponding closure relations.
The proofs are constructive and elementary: they rest entirely on the
classical closure relations for the Hermite and generalized Laguerre
functions, supplemented in three dimensions by the completeness of the
spherical harmonics.
No spectral theory beyond these standard results is required.

The present results complete the picture initiated by Szmytkowski and
Gruchowski~\cite{Szmytkowski2001} for the Dirac oscillator.
The structural comparison between the two cases is instructive.
For the DO the spinor nature of the eigenfunctions introduces an
off-diagonal closure sum $K$ whose vanishing requires the antisymmetry
$E_{-n} = -E_n$ of the spectrum, a non-trivial algebraic identity that
couples the two spinor components.
For the KGO no such complication arises: the squared equation is
scalar, the eigenfunctions reduce to Hermite or Laguerre functions
directly, and the closure sums are purely diagonal.
In this sense the KGO proof is not merely analogous to the DO proof
but strictly simpler, a reflection of the spin-$0$ nature of the
Klein--Gordon field.

A noteworthy feature of both the one- and three-dimensional proofs is
that the energy eigenvalues $E_n$ and $E_N$ play no role whatsoever.
Completeness is a property of the spatial eigenfunctions, which are
determined solely by the spatial part of the squared equation---itself
a non-relativistic harmonic oscillator equation in disguise.
The relativistic content enters only through the relation between $E$
and the oscillator quantum number, not through the shape of the
wavefunctions.
This observation has a practical consequence: the same closure
relations hold for \emph{any} value of $m$ and $\omega$, including the
massless limit $m \to 0$ and the ultra-relativistic regime
$E \gg m$.

Completeness is a prerequisite for several standard constructions that
have been carried out for the KGO in recent years but whose validity
implicitly required this result.
Among them:
\begin{itemize}
  \item \textbf{Propagators and Green's functions.}
        The energy-dependent Green's function of the KGO was derived
        using supersymmetric methods in Ref.~\cite{Junker2021}.
        The spectral representation of that Green's function,
        $G(\mathbf{r},\mathbf{r}';E)
         = \sum_{n_r,\ell,m}
           \Psi_{n_r\ell m}(\mathbf{r})
           \Psi_{n_r\ell m}^*(\mathbf{r}')/(E^2 - E_N^2)$,
        is rigorously justified only if the eigenfunctions form a
        complete set.

  \item \textbf{Thermal and statistical properties.}
        Partition functions and thermodynamic quantities for the KGO
        are typically expressed as sums over the complete set of
        energy eigenstates.
        Completeness ensures that no states are missed in such sums.

  \item \textbf{Perturbative expansions.}
        When the KGO is coupled to external fields, curved backgrounds,
        or deformed algebras~\cite{BoumaliMessai2014,BakkeFurtado2015,%
        MirzaNarimaniZare2011}, perturbation theory requires expanding
        the perturbed states in the unperturbed basis.
        The closure relation proved here provides the rigorous
        foundation for such expansions.
\end{itemize}

Several natural extensions of the present results suggest themselves.
First, the two-dimensional KGO---relevant for graphene-like systems
and planar quantum mechanics~\cite{BoumaliMessai2014}---has not been
treated here; its completeness proof would follow the same strategy,
with the Laguerre closure relation playing the central role and the
angular completeness supplied by the Fourier completeness of the
exponentials $e^{im\varphi}$.
Second, for the KGO in curved spacetimes or with topological defects,
the eigenfunctions are generally expressible in terms of confluent
Heun functions rather than Laguerre polynomials; completeness in those
settings is a more delicate question that would require appropriate
generalizations of the closure argument.
Third, the Dunkl--Klein--Gordon oscillator~\cite{HamilLutfuoglu2021},
whose eigenfunctions involve Laguerre and Jacobi polynomials with
Dunkl-deformed parameters, offers a natural next step within the
framework of the present approach.
We leave these extensions for future work.

We have proved that the eigenfunctions of the Klein--Gordon oscillator
form a complete orthonormal set in $L^2(\mathbb{R})$ (one dimension)
and $L^2(\mathbb{R}^3)$ (three dimensions).
The proofs are constructive, require no machinery beyond the classical
theory of orthogonal polynomials, and fill a gap that has been left
open in the literature since the introduction of the Klein--Gordon
oscillator by Bruce and Minning~\cite{BruceMinning1993}.

\bibliographystyle{unsrt}
\bibliography{Biblio}

\end{document}